\documentclass[aps,pre,twocolumn,amsmath,amssymb,superscriptaddress,showpacs,floatfix]{revtex4-1}

\usepackage{graphicx}
\usepackage{subfigure}
\begin{document}

\title{Percolation of sites not removed by a random walker in $d$ dimensions}
\author{Yacov Kantor}\email{kantor@tauex.tau.ac.il}
\affiliation{Raymond and Beverly Sackler School of Physics and
Astronomy, Tel Aviv University, Tel Aviv 69978, Israel}
\author{Mehran Kardar}
\affiliation{Department of Physics, Massachusetts Institute of Technology,
Cambridge, Massachusetts 02139, USA}
\date{\today}

\begin{abstract}
How does removal of sites by a random walk lead to blockage of
percolation? To study this problem of correlated site percolation, we
consider a random walk (RW) of $N=uL^d$ steps on a $d$-dimensional
hypercubic lattice of size $L^d$ (with periodic boundaries).
We systematically explore dependence of the probability $\Pi_d(L,u)$ of
percolation (existence of a spanning cluster) of sites {\em not removed}
by the RW on $L$ and $u$. The concentration of unvisited sites decays
exponentially with increasing $u$, while the visited sites are highly
correlated -- their correlations decaying with the distance $r$  as
$1/r^{d-2}$ (in $d>2$). On increasing $L$, the percolation probability
$\Pi_d(L,u)$ approaches a step function, jumping from 1 to 0 when $u$
crosses a percolation threshold $u_c$ that is close to 3 for all
$3\le d\le6$. Within numerical accuracy, the correlation length
associated with percolation diverges with exponents consistent with
$\nu=2/(d-2)$. There is {\em no percolation threshold} at the lower
critical dimension of $d=2$, with the percolation probability
approaching a smooth function $\Pi_2(\infty,u)>0$.

\end{abstract}
\maketitle
\makeatletter
\setlength{\@fptop}{0pt}
\makeatother
\section{Introduction}

In the simplest (Bernoulli) site or bond percolation
problem~\cite{Stauffer91,Gremmett99} sites or bonds of a regular
$d$-dimensional lattice are {\em independently} occupied with
probability $p$. For an infinite  system, there is a sharp percolation
transition point $p_c$, such that  for $p>p_c$ there exists an infinite
cluster spanning the system. Close to $p_c$, many geometrical and
physical properties become singular, as expressed by universal power-law
dependencies  on $|p-p_c|$. For example, the typical linear extension of
finite clusters of connected sites, indicated by the correlation length
$\xi$, diverges as $\xi\sim |p-p_c|^{-\nu}$. The {\em universal}
critical exponent $\nu$ depends only on space dimension $d$, and is well
known for Bernoulli percolation in all $d$ (see, e.g., Ref.~\cite{Koza16}). A much studied problem
in mathematical literature,  percolation has also been used to model a
variety of physical systems for the onset of connectivity and flow,
e.g., for current passing through a random resistor network.

An early application of percolation is to gels formed by random
crosslinking of polymers~\cite{Adam90}. {\em Gelation} has acquired new
interest in the context of reversible accumulations of nonspecific
biological molecules into liquid like droplets with important functions
as in transcription regulation~\cite{Sabari18}. The reverse process of
gel {\em degradation} is now also of relevance. In principle, the
removal of connections, rather than their addition, does not
qualitatively change the percolation picture. For example, in the process
of hydrogel degradation~\cite{Metters00,Metters00a,Metters01,Jahanmir18}
connections are severed mostly uniformly in space, and the measured
elastic and rheological properties~\cite{Diederich17,Schultz12} resemble
the gel formation process in reverse~\cite{Larsen08}.

While in Bernoulli percolation, the elements are added or removed randomly
and independently, new behavior emerges if subsequent events are
correlated. An extreme case is that of {\it explosive percolation} when
the events are specifically chosen so as to delay the percolation
transition~\cite{Raissa15}. A different extreme consists of removing entire
{\em straight lines} from the system in each step~\cite{K_PRB33}, e.g., by
drilling through a solid sample~\cite{Hilario15}, or by having very
elongated  obstacles (fibers) influencing molecular diffusion between
cells~\cite{Gomez19},  and also corresponds to a completely different
universality class~\cite{Hilario15,Schrenk16,Grassberger17a,Grassberger17}.
A variant of the latter is removal of sites or bonds performed by a
meandering {\em random walk} (RW).
This models a simplified version of a  degradation process in which a single  enzyme,
or possibly a few enzymes, travel through a gel, breaking the crosslinks
they encounter~\cite{Berry00,Fadda03}.
In an early numerical study of this problem, which they named ``random walk decay,"
Banavar {\em et al.}~\cite{Banavar85} considered properties of the clusters of
{\em vacant} sites, unvisited by the RW, on square and cubic lattices.
In a later study, Abete {\em et al.} considered percolation of the vacant bonds
on a cubic lattice~\cite{Abete04}, finding numerically the threshold and several critical
indices of the problem, which they called ``pacman percolation."

Independently, the mathematical community has also studied aspects of the above problem.
While physicists focused on geometry and critical properties of clusters, primarily in
$d=2$ and $d=3$ dimensions, mathematicians debated the very existence of a
percolation threshold. For a random walk of length $N$ to cover a finite fraction of the $M=L^d$
sites of a hypercubic lattice, the number of steps $N$ must be proportional to $M$.
The question of whether percolation of vacant sites stops for $N/M=u$ larger than a critical $u_c$
was addressed in several mathematical works, first for
$d\ge 7$~\cite{Sznitman08} and later for $d\ge3$~\cite{Sidoravicius09,Sznitman10}.
In the mathematical literature, this problem is referred to as ``percolation of vacant
sites of interlacement." Although, the bounds on $u_c$ established in
this literature were extremely broad, and in $d=3$ spanned many order
of magnitude~\cite{Balazs15}, we shall see that $u_c$ is approximately
3 for $3\le d\le 6$.

More precisely, in this work we perform a detailed numerical study of site percolation
on  $d$-dimensional  hypercubic lattices of linear dimensions $L$
($0\leq~\{x_1,x_2,\dots,x_d\}\leq~L-1$) and volume $V=a^dM=L^d$ (with lattice constant $a=1$).
A RW starts at an arbitrary initial position on the lattice and performs $N=uM=uL^d$ steps,
obeying periodic boundary conditions in all directions, i.e., $x_i=L$ coincides with
$x_i=0$ for $i=1,\dots,d$.
This scaling of $N$ maintains a fixed fraction of vacant sites in $d\ge 3$.
(The case of $d=2$ is discussed separately in Sec.~\ref{sec:perc2D}.)
We then define a configuration as ``spanning" (``percolating") if a continuous path of
vacant sites exists between boundaries at $x_d=0$ and $x_d=L-1$.
Note that while checking for this percolation condition, boundaries
in directions $1, \dots, d-1$ are assumed to be periodic;
this is sometimes referred to as ``helical boundary conditions"~\cite{Acharyya98}.
By considering a large number of realizations for
each $L$ and $u$, we determine the {\em spanning probability}
$\Pi_d(L,u)$. The fraction of vacant sites $p$ is a monotonically
decreasing function of $u$. In $d\ge 3$ there is a sharp
{\em percolation threshold} $u_c$, such that in the $L\to\infty$ limit
the spanning probability becomes a step function with $\Pi_d=1$ for
$u<u_c$  and $\Pi_d=0$ for $u>u_c$. (And in terms of $p$, this
corresponds to a threshold $p_c$.) The relation between $u$ and
$p$ for $d\ge 3$ is discussed in detail in Sec.~\ref{sec:VacantSites}.

Unlike Bernoulli percolation, the sites removed by a RW are highly correlated.
In Sec.~\ref{sec:corr_perc} we demonstrate that the correlations between vacant sites
decay as a power-law, proportional to $1/r^{d-2}$ with their separation $r$.
Such correlated percolation has been argued to belong to a universality
class characterized by an exponent $\nu=2/(d-2)$ for divergence of
the correlation length~\cite{Weinrib83,Weinrib84}.
We  set up to test this behavior in dimensions $3\leq d\leq6$.
Section~\ref{sec:perc3D} details our numerical study of  percolation in $d=3$,
while Sec.~\ref{sec:percHighD} extends the exploration to $d=4$, 5, and 6.
There is no percolation threshold in $d=2$, resulting in a smooth limiting function
for spanning probability as demonstrated in Sec.~\ref{sec:perc2D}.
The limits of this function for very short and very long walks are
discussed in Sec.~\ref{sec:ShortAndLong}.
We conclude in Sec.~\ref{sec:discussion} with some topics for future investigation.

\section{Statistics of sites not visited by a random walk}\label{sec:VacantSites}

\begin{figure}[t!]
\includegraphics[width=8 truecm,  clip=true]{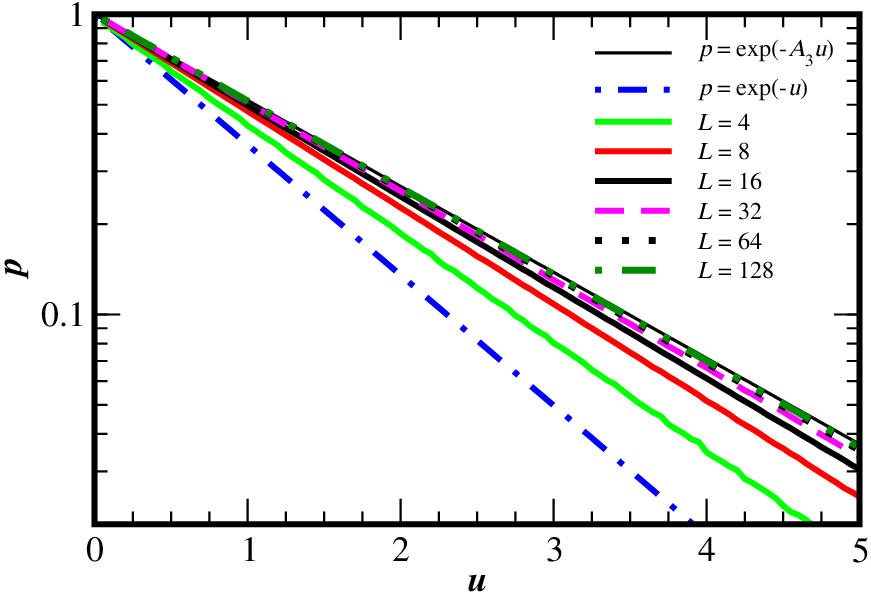}
\caption{Semilogarithmic plot of the probability (fraction) $p$ of vacant
sites as a function of $u$ at $d=3$. The dash-dotted bottom
curve corresponds to $p={\rm e}^{-u}$ obtained by dropping
$uL^3$ independent sites on a lattice of $L^3$ sites. The topmost line represents the
asymptotic behavior $p=\exp(-A_3u)$. The remainder of the curves represent
measured values of $p$ as functions of $u$ for $L=4, 8,\dots,128$ (bottom
to top). Data points are spaced $\Delta u=0.05$,
and each point is an average of $400$ configurations.
}
\label{fig:PvacantVsu}
\end{figure}

As a random walker performs $N=uL^d$ steps on a finite hypercubic
lattice of volume $M=L^d$ (``box") with periodic boundary conditions, the
fraction of {\em unvisited} or {\em vacant} sites $p$ decreases with increasing
$u$. If instead of performing an $N$-step RW, we randomly and independently
select $N$ of sites on a lattice, the occupied sites would be random,
uncorrelated, with some of them multiply occupied.
More precisely, the fraction of vacant sites would be   $p=\exp(-u)$,
as indicated by the bottom dash-dotted line in Fig.~\ref{fig:PvacantVsu},
with their positions uncorrelated as in usual (Bernoulli) percolation.
However, the obvious correlations in the positions of the random walker
generate a different dependence $p(u)$ as $L\to\infty$, which also remains nonzero
for any finite $u$.

The fractal dimension of a RW is 2, leading to qualitatively different behaviors
in $d=2$ that are separately discussed in Sec.~\ref{sec:perc2D}.
The results described in this section, as well as in Sec.~\ref{sec:corr_perc},
thus pertain only to dimensions $d\geq3$.
While the RW on an {\em infinite} hypercubic lattice does intersect itself,
it can be shown rigorously that the number of {\em distinct} visited sites of a long walk
of $N$ steps grows as~\cite{Vineyard63,Montroll65,Rubin82}
\begin{equation}\label{eq:Ndist}
N_{\rm dist}=A_dN\,,
\end{equation}
with subleading corrections of $O(N^{1/2})$ in $d=3$, and $O(\ln N)$
in $d=4$. The corrections do not increase with $N$ for $d\ge 5$,
as self-intersections of remote parts of RWs become negligible.
The coefficient $A_d$ in Eq.~\eqref{eq:Ndist} depends on lattice
type and space dimension. It is the inverse of the mean number
$B_d$ of visits of a random walker to its initial position~\cite{Aldous83,Nemirovsky90,Brummelhuis91}
(see also chapter 3 in Ref.~\cite{hughes_book}).
[Also, the mean number of RW steps required
to visit {\em all} sites of a {\em finite box} of size $M$ is
$B_dM\ln M$~\cite{Brummelhuis91} (see also Ref.~\cite{Weiss85}).]
A detailed procedure for calculation of both $A_d$ and subleading
corrections for various lattices was outlined by  Montroll and
Weiss~\cite{Montroll65}, while an efficient numerical method can
be found in Ref.~\cite{Griffin90}. The values of these coefficients for
hypercubic lattices are $A_3=0.659$~\cite{Montroll65}, $A_4=0.807$,
$A_5=0.865$ and $A_6=0.895$~\cite{Griffin90}. (Here and thereafter,
the accuracy of numbers without error bars is a single unit of the
last digit or better.)

Sites visited by a RW are strongly correlated. In an infinite space,
an $n$-step RW explores a volume of radius $r\sim n^{1/2}$ in which, for
$d\ge3$, the density of visited sites will be $n/r^d\sim~1/r^{d-2}$.
Thus, if a particular lattice site on a hypercubic lattice is visited by
a RW, then the probability of finding another visited site at a large separation
$r$ will be $C_d/r^{d-2}$, where the lattice- and
dimension-specific prefactor $C_d$ is of order of unity. Consider a random
variable $w(\vec{x})$ which is 1, if a site at position $\vec{x}$ has
been visited by a RW, and is 0 otherwise. Probabilities, of RW visiting
positions $\vec{x}$ and $\vec{y}$ are correlated, such that
$\langle w(\vec{x})w(\vec{y})\rangle/\langle w(\vec{x})
\rangle\approx C_d/|\vec{x}-\vec{y}|^{d-2}$, where $\langle\rangle$ denotes
an average over realizations of RWs.

To examine the problem on a finite hypercube with periodic boundary
conditions, we  first create a RW on an infinite lattice, then
tile the space with boxes (hypercubes) of size $M=L^d$.
The boxes are then cut out and superposed; a procedure that
we refer to as ``folding."
For $d\ge3$, and
for finite $u=O(1)$  and large $L$, the walk on an infinite lattice
will  have typical size (end-to-end distance) $\sim L^{d/2}\gg L$. In
an infinite space most of the boxes will be empty, and the RW will visit
of order $L^{d-2}$ distinct boxes. In the $i$th visited cube there will
be $n_i$ {distinct} visited sites, and $n_i$ will be of order $L^2$.
Thus, the fraction of {\em unvisited} ({\em vacant}) sites in a cube is $p_i=1-n_i/L^d\simeq\exp(-n_i/L^d)$, as $n_i/L^d\sim 1/L^{d-2}$
is very small. The ``folding" process from infinite space into a single
periodic box effectively removes correlations between far away boxes,
and we can treat the configurations arising from superposition of
distinct boxes as uncorrelated. Therefore, the probability that a
particular point in the periodic box has not been visited is
\begin{eqnarray}\label{eq:pu}
p=\prod_ip_i=\exp\left(-\sum_i\frac{n_i}{L^d}\right)\nonumber\\
=\exp\left(-\frac{N_{\rm dist}}{L^d}\right)=\exp(-A_du)\,.
\end{eqnarray}
This asymptotic (very large $L$) result has been rigorously proven
in Ref.~\cite{Brummelhuis91}.

In numerical simulations with moderate values of $L$ ($\sim$10--100),
 the finite-$N$ corrections to  Eq.~\eqref{eq:Ndist} are noticeable, especially in $d=3$
 where there is a non-negligible probability for a RW that exited a box
(on the infinite lattice) to return to it. Yet correlations of occupied
sites even in adjacent
boxes do not exceed $C_d/L^{d-2}$, and are much smaller for nonadjacent
boxes (which are the majority). Thus, in the ``folding" process we
superimpose practically independent configurations. Also, the
fraction of visited sites  $n_i/L^d\sim1/L^{d-2}$ is still significantly
smaller than 1.
The deviations from Eq.~\eqref{eq:pu} for moderate $L$
can be corrected for by replacing $A_d$ with an effective $A_d(L)$.
We examined numerically the functions $p(u)$ in $d=3$, 4, 5, and 6,
and in all cases the results for various $L$ could be fitted
extremely well by a pure exponential $p=\exp[-A_d(L)u]$. For $d=3$
we checked such dependence for $L=4$, 8,\dots,128. The results presented
in Fig.~\ref{fig:PvacantVsu} fit pure exponentials at the accuracy
level of $\chi^2\sim 10^{-7}$. The slopes on the semilogrithmic
plots depend on $L$ but converge very fast to the known value of $A_3$
indicated by the top line on Fig.~\ref{fig:PvacantVsu}.
Of course, both this and the following statements, about ``purely
exponential behavior," should not be taken in a strict mathematical
sense: We know, that for a {\em finite} periodic box of size $M$, a walk
of $B_dM\ln M$ steps will ``completely occupy" the box. So, in $d=3$ and
$L=128$, this will occur for $u>B_3\ln 128^3\approx 22$, far beyond the limits
of applicability of the above discussion.

\begin{figure}[t]
\includegraphics[width=8 truecm, clip=true]{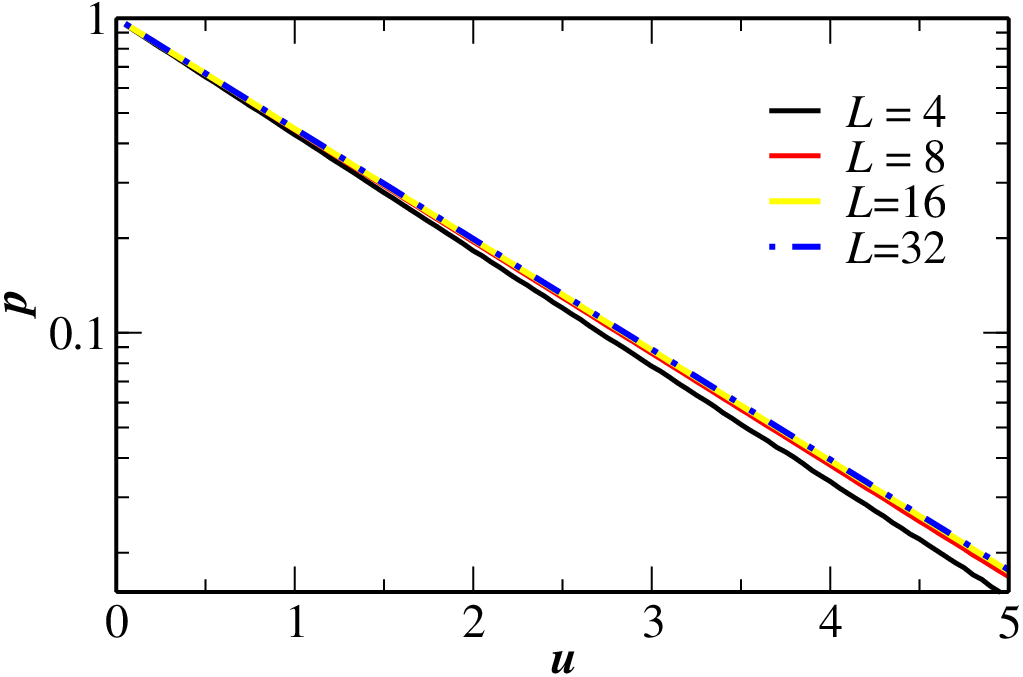}
\caption{
Semilogarithmic plot of the fraction $p$ of unvisited sites as a function
of parameter $u$ controlling the length of the random walk in $d=4$, for
$L=4$, 8, 16, and 32 (bottom to top). Data points are spaced $\Delta u=0.05$,
and each point is obtained from $10^4$ configurations.
}
\label{fig:4D_p_vs_u}
\end{figure}

In $d\ge 4$ the distant parts of RW rarely intersect and therefore
$A_d(L)$ converges very fast to its asymptotic value $A_d$. As an
example, in Fig.~\ref{fig:4D_p_vs_u} we show the $d=4$ case, where
for lattices of sizes $L=4, 8, 16$, and 32 we generate $10^4$
configurations per data point to measure the relation between $u$
controlling the chain length and the fraction of unvisited sites $p$.
As before, all curves are straight on semi-log scale, i.e., $p$ is an
exponentially decaying function of $u$. The rate of that decay quickly
converges to a constant, leading to $p=\exp(-A_4u)$ for large $L$.
We repeated these calculations also for $d=5$ and $d=6$. There was
practically no $L$ dependence of $A_d(L)$ and the coefficients
reached their asymptotic values already for small $L$.

\section{Correlated percolation}\label{sec:corr_perc}

To quantify the correlations between sites visited by the RW following
the ``folding" from infinite space, let us examine the two-point correlation function
$\langle w(\vec{x})w(\vec{y})\rangle$.
We argued previously that the RW segments from distinct boxes are almost
uncorrelated, after being ``folded" into the single box with periodic boundary conditions.
A randomly selected site $\vec{x}$ in the periodic system has a probability
$\langle w(\vec{x})\rangle=1-p=1-\exp(-A_du)$ of being occupied.
This point may have been visited a few times either by the same RW segment,
or by RW segments belonging to distinct boxes before ``folding."
Either way, the probabilities of finding
a site at position $\vec{y}$ that has been visited by the segment of RW in
{\em the same} box (before ``folding") is given by
$C_d/|\vec{x}-\vec{y}|^{d-2}$, and it is modified by the number of distinct boxes
to which the segments belonged before ``folding." Thus, the correlated
part of $\langle w(\vec{x})w(\vec{y})\rangle$ will decay as
$1/|\vec{x}-\vec{y}|^{d-2}$, as long as $|\vec{x}-\vec{y}|$ is significantly
smaller than the box size $L$. Of course, the chances that position $\vec{y}$
also has been visited is dominated by segments of RW that belong to
{\em other} boxes before the ``folding." This constant background part
must be subtracted when measuring the cumulant, leading to
$\langle w(\vec{x})w(\vec{y})\rangle_c=F(u)/|\vec{x}-\vec{y}|^{d-2}$, where
$F(u)$ is a smooth function of order unity for $u=O(1)$.
(In $d=3$, an equivalent argument can be found in Ref.~\cite{Fadda03}.)

Since we are interested in the properties of {\em vacant} sites, we
define a related variable $v(\vec{x})=1-w(\vec{x})$, with $\langle v(\vec{x})\rangle\equiv p$.
Products of these
variables at distinct positions have the property
$\langle v(\vec{x})[v(\vec{y})+w(\vec{y})]\rangle=\langle v(\vec{x})\rangle=p$
and
$\langle w(\vec{x})[v(\vec{y})+w(\vec{y})]\rangle=\langle w(\vec{x})\rangle=1-p$.
Since our system is translationally invariant and symmetric under inversion, we
have $\langle v(\vec{x})w(\vec{y})\rangle=\langle w(\vec{x})v(\vec{y})\rangle$.
Thus, by subtracting the two previous equalities from each other, we find
$\langle v(\vec{x})v(\vec{y})\rangle=\langle w(\vec{x})w(\vec{y})\rangle+2p-1$.
Consequently, the vacant sites exhibit the same decay of cumulants as  visited sites, i.e.,
\begin{equation}\label{eq:cumul}
\langle v(\vec{x})v(\vec{y})\rangle_c\sim 1/|\vec{x}-\vec{y}|^{d-2}.
\end{equation}

The two-point correlation function captures only one aspect of the interesting
information about the system. The RW in a periodic box has the distinct
property of being a {\em single} cluster (noting periodic
boundary conditions). While vacant sites can form many clusters, they
tend to aggregate into one large cluster. The unusual properties
of both small and ``infinite" clusters have been studied by several
authors~\cite{Banavar85,Teixeira09,Teixeira11,Benjamini08,Abete04}.

In ``usual" (Bernoulli) percolation there are no correlations
between occupied sites or bonds.
This problem has a lower critical dimension of
$d=1$, where $p_c=1$, and an upper critical dimension of $d_c=6$~\cite{Toulouse74},
above which mean-field behavior is expected; e.g., with the correlation length
diverging  with exponent $\nu_{\rm B}=1/2$. Weinrib analyzed stability of the 
Bernoulli percolation universality class to correlations~\cite{Weinrib84}
(following a similar treatment for critical phase transitions~\cite{Weinrib83}).
By appealing to a generalized Harris criterion~\cite{Harris74}, he demonstrated 
that short-range correlations, as well as power-law correlations decaying 
as $1/r^a$ with $a>d$, do not modify the universality class of Bernoulli
percolation. However, for $a<d$, the relevance of the correlations
is determined by the extended Harris criterion~\cite{Weinrib83}:
If $a\nu_{\rm B}-2<0$, then the correlations are relevant. Equation~\eqref{eq:cumul}
shows that vacant sites have a power law correlation
with $a=d-2$. The quantity $(d-2)\nu_{\rm B}-2$ is $-1.12$, $-0.62$,
$-0.29$ and 0, for $d=3, 4, 5$, and 6, respectively~\cite{Koza16}.
This expression becomes positive for $d\ge 7$.

Percolation with long-range correlations is a well-researched
subject~\cite{Coniglio09}. (For a more recent study see Ref.~\cite{Gori17}.)
In most physical contexts, correlations are generated by thermodynamic
systems such a critical Ising model. It has been
shown~\cite{Weinrib83,Weinrib84} that, for power-law correlations, the
exponent characterizing divergence of the correlation length equals
$\nu=2/a$, which in our case is
\begin{equation}\label{eq:nu}
\nu=2/(d-2),\ \ {\rm for}\ \ 3\le d\le 6.
\end{equation}
Abete {\em et al.}~\cite{Abete04} studied percolation of vacant
{\em bonds} on lattices of up to $60^3$ and found $\nu=1.8\pm0.1$,
consistent with $\nu=2$ expected from Eq.~\eqref{eq:nu}. In this
work we consider {\em site} percolation on large lattices, and
confirm Eq.~\eqref{eq:nu} in $d=3, 4$, and 5. In $d=d_c=6$  power-law
correlations $\sim 1/r^4$ represent a {\em marginal} perturbation,
and both Eq.~\eqref{eq:nu} and standard Bernoulli percolation lead to
$\nu=1/2$.

In a system of infinite size the spanning probability is a step
function, jumping between 1 and 0 as the percolation threshold
is crossed. However, for finite $L$ it exhibits a smooth crossover
between these values. It has been shown for Bernoulli
percolation~\cite{Langlands92,Cardy92,Langlands94,Watts96} that for
large $L$ the value of spanning probability $\Pi_{d,{\rm B}}(L,p)$
at $p=p_c$ reaches a universal value  independent of {\em micro}scopic
lattice details or the consideration of bond or site percolation.
Thus, the {\em critical spanning probability}
$\Pi^c_{d,{\rm B}}\equiv\lim_{L\to\infty}\Pi_{d,{\rm B}}(L,p_c)$ is
a universal number characterizing a percolation universality class.
For example, $\Pi^c_{2,{\rm B}}=1/2$ in two-dimensional Bernoulli
site and bond percolation with free boundary conditions. However,
the value of the spanning probability  for finite $L$,
and, consequently, its universal limit of $\Pi^c_{d,{\rm B}}$,
{\em does} depend on the {\em macro}scopic definition of spanning,
such as requiring spanning in several directions simultaneously or
using periodic {\em versus} free boundary conditions. Such dependence
on macroscopic definitions has been observed by several authors~\cite{Stauffer94,Hovi96,Lin98,Acharyya98}.
The value of $\Pi^c$ thus may serve as an extra indicator of differences
between universality classes. In the problem of percolation of vacant
sites of a RW, we can similarly define
$\Pi_d^c\equiv\lim_{L\to\infty}\Pi_d(L,u=u_c)$ characterizing this
type of percolation.

\section{Percolation in $d=3$}\label{sec:perc3D}

\begin{figure}[h]
\includegraphics[width=8 truecm, clip=true]{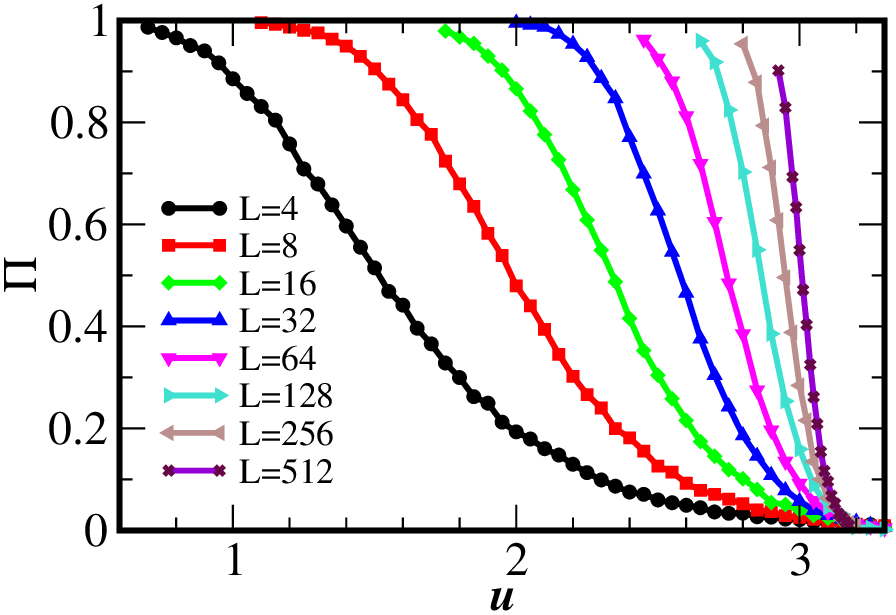}
\caption{
Percolation (spanning) probability $\Pi_3$ of vacant sites on a cubic lattice
in $d=3$ as a function of $u$ for $L=4,8,\dots,512$ (left to right).
}
\label{fig:percVsu}
\end{figure}

In $d=3$ we considered site percolation along the $x_3$ direction for cubic
lattice sizes  $L=4,8,16,\dots,512$. For every $L$ and $u$ we
generated many realizations of RWs of $N=uL^3$ steps. Spanning
probability $\Pi_3(L,u)$ was calculated by averaging $10^4$
realizations for each $u$ for $L\le64$, and 4000 configurations for
each $u$ for $L\ge 128$. The limiting factor in the computations
was the large lattice size, and correspondingly long walks reaching
$N=5\times 10^5$, necessitating long times required to process each
configuration. (The previous study by Abete {\em et al.}~\cite{Abete04}
reached lattice
sizes $L=60$ and considered bond percolation.) Figure~\ref{fig:percVsu}
depicts the spanning probability as a function of $u$. As expected, the
transition becomes sharper with increasing $L$. The curves also exhibit
a very strong drift toward larger values of $u$ with increasing $L$.

\begin{figure}[h]
\includegraphics[width=8 truecm, clip=true]{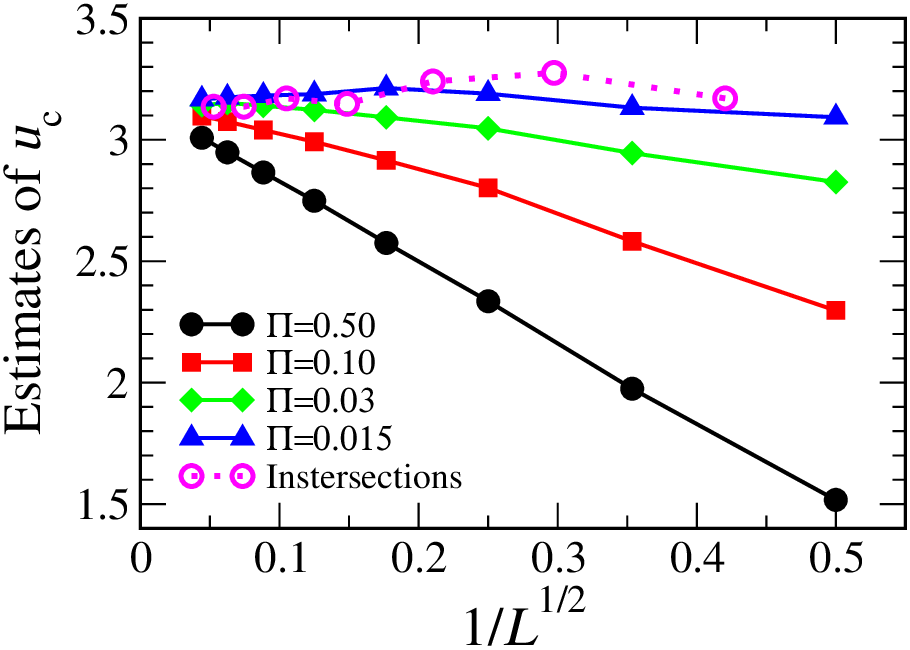}
\caption{
Successive estimates $u_*$ of  the percolation threshold $u_c$, as a function of $1/\sqrt{L}$ for
$L=4,8,\dots,512$ (full circles and solid lines). Different curves provide the estimates obtained from
numerical values of $u_*$ for which the percolation probability is $\Pi_3(L,u_*)=0.5$, 0.1, 0.03 or 0.015 (bottom to top).
An additional estimate of $u_c$ (open circles and dotted line) is obtained from intersection of
the spanning curve for a particular $L$ with the corresponding curve for $2L$; the
resulting values are plotted as a  function of
$1/\sqrt{L_{\rm eff}}\equiv 1/(2^{1/4}\sqrt{L})$.
}
\label{fig:Threshold}
\end{figure}

There are several methods to determine the
critical value $u_c$ for percolation. The low accuracy of the results
prevents us from using points $u=u_*$ of maximal slope of $\Pi_d(L,u)$
as estimates of $u_c$. However, we can examine the $L$-dependence of
 values $u_*$ for which $\Pi_d(L,u_*)=c$. Independently of the
choice of $c$, we expect $\lim_{L\to\infty}u_*=u_c$. Moreover, for
large $L$, we expect
\begin{equation}\label{eq:Delta_u}
|u_*-u_c|\sim L^{-1/\nu}.
\end{equation}
For $d=3$, Fig.~\ref{fig:Threshold} depicts the dependence of
successive estimates of $u_c$ on $L^{-1/2}$, as suggested by
Eq.~\eqref{eq:Delta_u} with anticipated $\nu\approx 2$. (Note that
for small $L$, the estimates of $u_c$ do not follow the asymptotic 
form of Eq.~\eqref{eq:Delta_u}, and are sometimes even  nonmonotonic
functions  of $L$.) All four lines extrapolate to $u_c=3.15\pm0.01$.

An additional set of  estimates (open circles) is obtained by looking at
{\em points  of intersection} $u_*$ of two sequential curves of $\Pi_d$:
E.g., for $L_1$ and $L_2$ we may look for $\Pi_d(L_1,u_*)=\Pi_d(L_2,u_*)$
and study the resulting $u_*$ as a function of $(L_1L_2)^{-1/2\nu}$. For $d=3$,
when $L_1=L$, then $L_2=2L$. This sequence of estimates leads to the
same $u_c$. This $u_c$ corresponds to $p_c=0.125\pm0.001$, which is rather
close to the threshold of $p_c=0.139\pm0.001$  found for
{\em bond} percolation in Ref.~\cite{Abete04}. The fact that the site percolation
threshold found in our work is {\em smaller} than the bond percolation
threshold of Ref.~\cite{Abete04} is somewhat surprising.

By following the value of $\Pi_3(L,u_*)$, at points $u_*$ where curves
$\Pi_3(L,u)$ for successive $L$ intersect, we estimate the critical
spanning probability $\Pi^c_3=0.04\pm0.01$. This value is slightly larger
than $\Pi^c_3\approx0.032$ found by Abete {\it et al.}~\cite{Abete04}
in bond percolation; both are significantly smaller
than the analogous number 0.513~\cite{Acharyya98} in Bernoulli
percolation.

Close to the percolation threshold, the correlation length $\xi$ diverges
as $|u-u_c|^{-\nu}$. For finite system size $L$, as long as
$L>\xi$ the percolation probability resembles that of an infinite system,
i.e., $\Pi_d(L,u)\approx1$ or $\approx 0$, for $u<u_c$ or $u>u_c$,
respectively. For $\xi\gtrsim L$ the value of $\Pi_d(L,u)$ decreases
from close to 1 to near 0 as $u$ increases. Therefore, the
transition region approximately  appears when
$b|u-u_c|^{-\nu}> L$, where $b$ is a numerical prefactor. Thus, the
width of the transition region scales as $\Delta u\approx (L/b)^{-1/\nu}$.
Since $\Pi_d$ changes between 1 and 0 in the transition region, we
expect the absolute value of the slope in that region
to be $\approx 1/\Delta u$ or
\begin{equation}\label{eq:slope}
{\rm slope}\approx (L/b)^{1/\nu}.
\end{equation}
(Here and thereafter we disregard the negative sign of the slope.)

\begin{figure}[t!]
\includegraphics[width=8 truecm, clip=true]{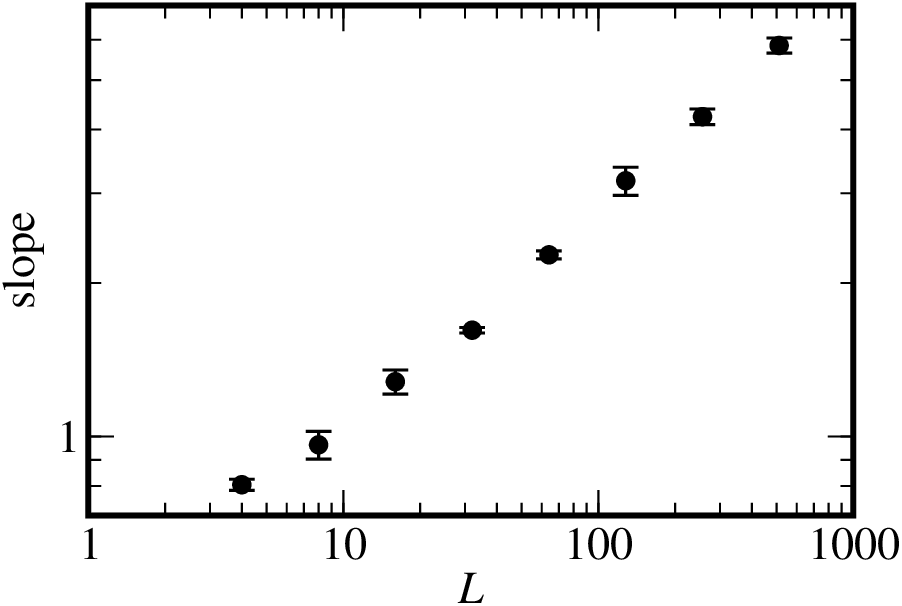}
\caption{
Logarithmic plot of the inverse width of the percolation transition (in terms of the
variable $u$) as a function of $L$. $1/\Delta u$ is measured from the maximal slope
of the curves in Fig.~\ref{fig:percVsu}. In theses coordinates, it occurs approximately
where the percolation probability is 1/2.
}
\label{fig:Correlation}
\end{figure}

\begin{figure*}[t!]
     \subfigure[\label{fig:1}]{%
       \includegraphics[width=8 truecm]{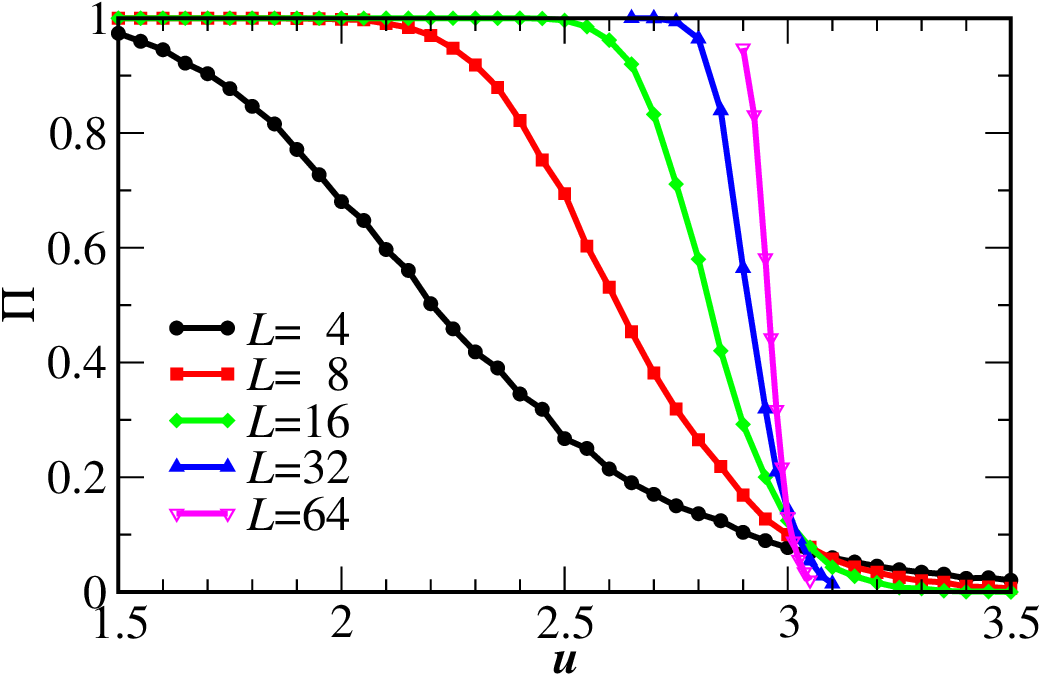}
     }
    \hfill
     \subfigure[\label{Fig:2}]{%
       \includegraphics[width=8 truecm]{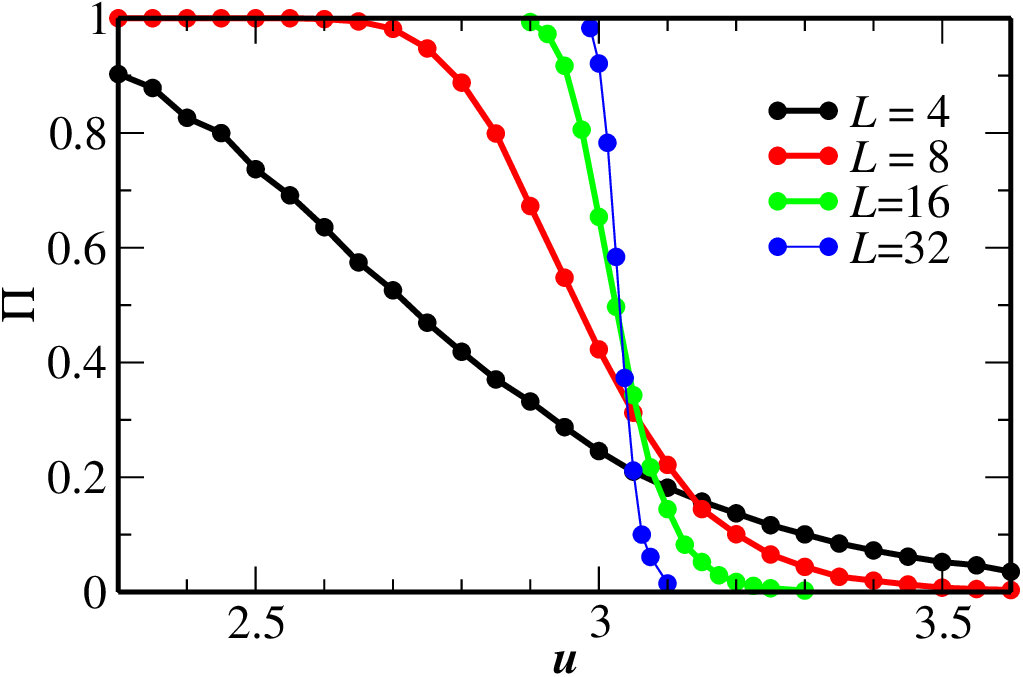}
     }
\caption{Spanning probabilities for (a) $d=4$, and (b) $d=5$. The transitions
become steeper with increasing $L$. Each data point for large $L$ is an
average over 4000 samples. For smaller $L$ [$\le 8$ in (a), and $\le 16$ in (b)]
$10^4$ configurations were generated.}
     \label{fig:4Dand5D_vs_u}
\end{figure*}

In $d=3$ we measured the absolute value of the maximal slope of the
curves in Fig.~\ref{fig:percVsu} for each $L$, and the results are
presented in Fig.~\ref{fig:Correlation}. According to Eq.~\eqref{eq:slope},
the slope of this relation on a logarithmic plot should be $1/\nu$. We
observe a slow and noisy increase of the slope with increasing $L$.
The inverse of the extrapolated  slope produces an estimate of
$\nu=2.04\pm 0.08$ that is larger than $\nu=1.8\pm0.1$ that was
found on significantly smaller systems in Ref.~\cite{Abete04}
and agrees well with the value $\nu=2$ predicted by Eq.~\eqref{eq:nu}.
Instead of using the largest slopes for each $L$,
we could have concentrated on significantly smaller slopes of
the curves at $u=u_c$. Such an approach may provide a useful estimate
of possible systematic errors. However, our data are too noisy and inaccurate
in this area to produce reliable results.

\section{Percolation in $d=4$, 5, and 6}\label{sec:percHighD}

In higher dimensions $d$, we followed a similar strategy to the one
described in the previous section. For $d=4$ and 5, we generated RWs
of $N=uL^d$ steps starting from $L=4$ and then doubling $L$
until  $L=64$ for $d=4$ or $L=32$ for $d=5$. (With increased $d$,
we had to limit the maximal size $L$.) For each $L$, from repeated
tests of spanning along $x_d$, we determined $\Pi_d(L,u)$. Results
of these measurements are depicted in Fig.~\ref{fig:4Dand5D_vs_u}.

The ``drift" of the curves to the right in $d=4$, and especially
in $d=5$, is significantly smaller than in $d=3$,  allowing
a rather accurate determination of the percolation threshold.
In $d=4$ we find $u_c=2.99\pm0.01$, which corresponds to occupation fraction
$p_c=0.0898\pm0.0007$ of unvisited sites. This
value is roughly half of the percolation threshold of 0.197 for regular
(Bernoulli) site percolation on a hypercubic lattice in
$d=4$~\cite{Mertens18}. In $d=5$  we  estimate $u_c=3.025\pm0.008$,
which corresponds to the critical fraction of vacant sites
$p_c=0.0730\pm0.0006$. This result is again about half of the
percolation threshold of 0.141 for Bernoulli site percolation
on a hypercubic lattice in $d=5$~\cite{Koza16}. It is difficult
to determine the critical spanning probability $\Pi_d^c$ with
any accuracy. However, by following the values of the intersections
of $\Pi_d(L,u)$ curves with sequential values of $L$, we estimate
that $\Pi_4^c$ is between 0.1 and 0.2, while $\Pi_5^c$ is about 0.4.

We next estimate values for the critical exponent $\nu$
from the dependence  of slope  on $L$ near the transition point.
In $d=4$, we were able to measure accurately both the maximal
slope and the slope at the estimated $u_c$. Differences between
these two methods provide us with an estimate of the possible
systematic error. We find that $\nu=1.0\pm0.1$, which agrees with
$\nu=1$ expected from Eq.~\eqref{eq:nu}, and is very different
from the Bernoulli value of $\nu_{\rm B}=0.685$~\cite{Koza16}. In $d=5$
the largest slope practically coincides with the slope at $u_c$
(except for $L=4$). We detect a slight dependence of the effective
exponent $\nu_{\rm eff}$ on $L$, and the extrapolated value is
$\nu=0.65\pm0.03$, where the error bars reflect the uncertainty in
the extrapolation procedure. This agrees excellently with $\nu=2/3$
expected from Eq.~\eqref{eq:nu}, and differs from the regular
(Bernoulli) percolation exponent of $\nu_{\rm B}=0.57$~\cite{Koza16}.

\begin{figure}[t]
\includegraphics[width=8 truecm, clip=true]{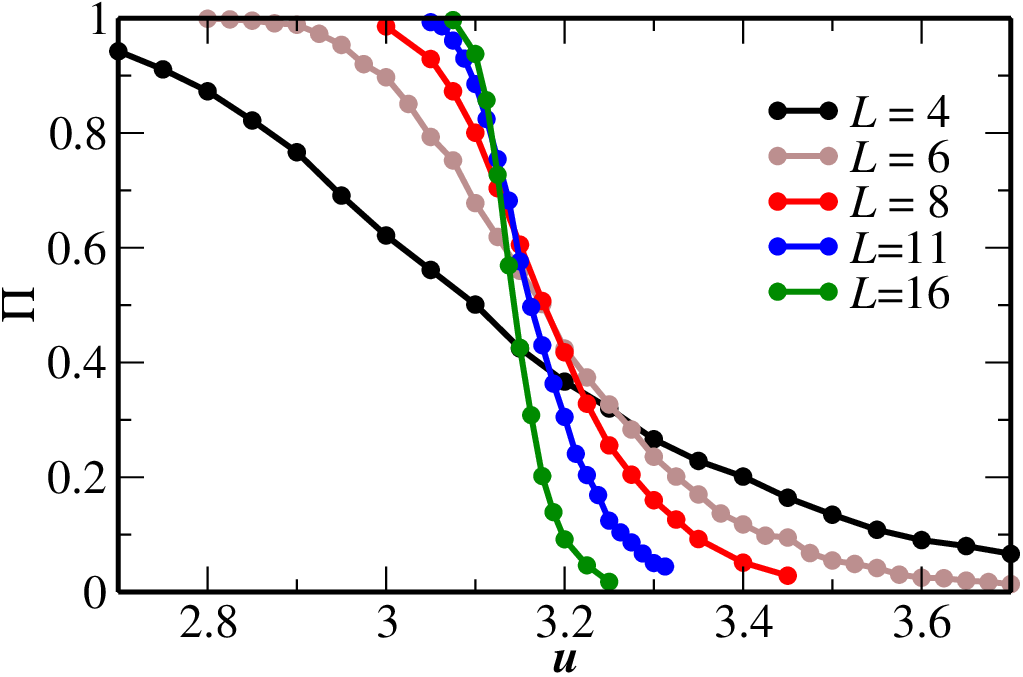}
\caption{
Spanning probability $\Pi_6$, for $d=6$, as a function of  $u$, for
$L=4$, 6, 8, 11, and 16 (from gradual to steep decrease). For $L=4$ each data
point corresponds to $10^4$ configurations, while for $L\ge6$, 4,000
configurations were sampled.
}
\label{fig:6D_Pi_vs_u}
\end{figure}

In $d=6$, we generated random walks of length $uL^6$ for $L=4, 6, 8, 11$,
and 16. This required dealing with lattice sizes $M\approx1.7\times 10^7$
and RWs reaching $6\times 10^7$ steps. Figure \ref{fig:6D_Pi_vs_u} depicts
the spanning probability $\Pi_6(L,u)$ as a function of $u$ for several
values of $L$. The maximal rate of decrease of all the curves is close to
the point where $\Pi\sim 1/2$ or slightly higher. There is almost no drift
in the curves with increasing $L$, and we estimate $u_c=3.10\pm0.05$, which
corresponds to critical fraction of vacant sites $p_c=0.062\pm0.003$. The
latter is significantly smaller than the threshold 0.109 for
Bernoulli site percolation~\cite{Koza16}. The intersection points between
$\Pi_6(L,u)$ for successive $L$ drift strongly upward, leading to a rather
large estimate of $\Pi_6^c\sim0.9$ for the critical spanning probability.
Since $d=d_c=6$ is the common upper critical dimension~\cite{Toulouse74},
both  the regular percolation theory and Eq.~\eqref{eq:nu} posit $\nu=1/2$.
Unfortunately, the maximal lattice size of $L=16$ computationally available
to us, is  too small to determine $\nu$ with any degree of accuracy.
In addition to the obvious limitations of small $L$, we note that at $d_c$
there are logarithmic corrections in cluster size
distributions~\cite{Essam78,Nakanishi80} that further complicate detection
of the trends. A straightforward power-law fit in the range $4\le L \le16$
produces $\nu_{\rm eff}\approx 0.6$. Due to the above mentioned limitations,
we believe that our numerical result does not contradict the expected value
of $\nu=1/2$.

\section{Spanning probability in $d=2$}\label{sec:perc2D}

The behavior of a two-dimensional RW of $N=uL^2$ steps on
the $L\times L$ square lattice is quite different from the higher
dimensions discussed previously. For $u=O(1)$, the root mean-squared
end-to-end distance of a walk is of the order of the linear size $L$,
rendering the imagery of multiple ``foldings" of a much longer RW  in
Sec.~\ref{sec:VacantSites} inapplicable. As in higher dimensions, we will
test for spanning along the $x_2$ direction, while assuming periodicity in
the $x_1$  direction. Clearly, for $u\ll1$ a typical walk is simply too
short to block percolation of vacant sites along $x_2$, while for $u\gg1$
a single ``circumnavigation" of the walk in $x_1$ direction will almost
certainly block the percolation in $x_2$  direction. In earlier work,
Banavar {\em et al.}~\cite{Banavar85} studied properties of clusters of
vacant sites in a system of this type. They  were primarily interested in
the fractal and fracton dimensions in the regime where the RW covers a
finite fraction of a lattice, i.e., for $u$  of order of unity.

Since the fractal dimension of a RW is 2, it is not surprising that its
behavior in $d=2$ exhibits  important differences from higher dimensions.
For example, the number of {\em distinct} sites visited by an $N$-step
walk in $d=2$ on an {\em infinite} lattice is modified from
Eq.~\eqref{eq:Ndist} to the leading order
as~\cite{Dvoretzky51,Erdos60,Vineyard63,Montroll65}
\begin{equation}\label{eq:Ndist2D}
N_{\rm dist}=A_2 N,\ \ {\rm with}\ \ A_2=\pi/\ln N.
\end{equation}
Correspondingly, the number of visits of a random walker to its initial
position increases logarithmically with $N$, namely $B_2=(\ln N)/\pi$,
as opposed to a constant $B_d$ for $d>2$. Similarly, the number of steps
required to visit all sites on a lattice of $M$ sites is $\sim M\ln^2M$,
i.e. with an extra logarithm compared to higher dimensions. Thus, when
considering RWs of $N=uL^2$ steps on a square lattice of  $M=L^2$ sites,
with $u=O(1)$, the density of occupied sites actually vanishes in the
limit $L\to\infty$. This again justifies arguments used in
Sec.~\ref{sec:VacantSites} to demonstrate that, even for moderate $L$,
there is a pure exponential dependence of the fraction of vacant sites
$p$ on the parameter $u$.

\begin{figure}[t]
\includegraphics[width=8 truecm, clip=true]{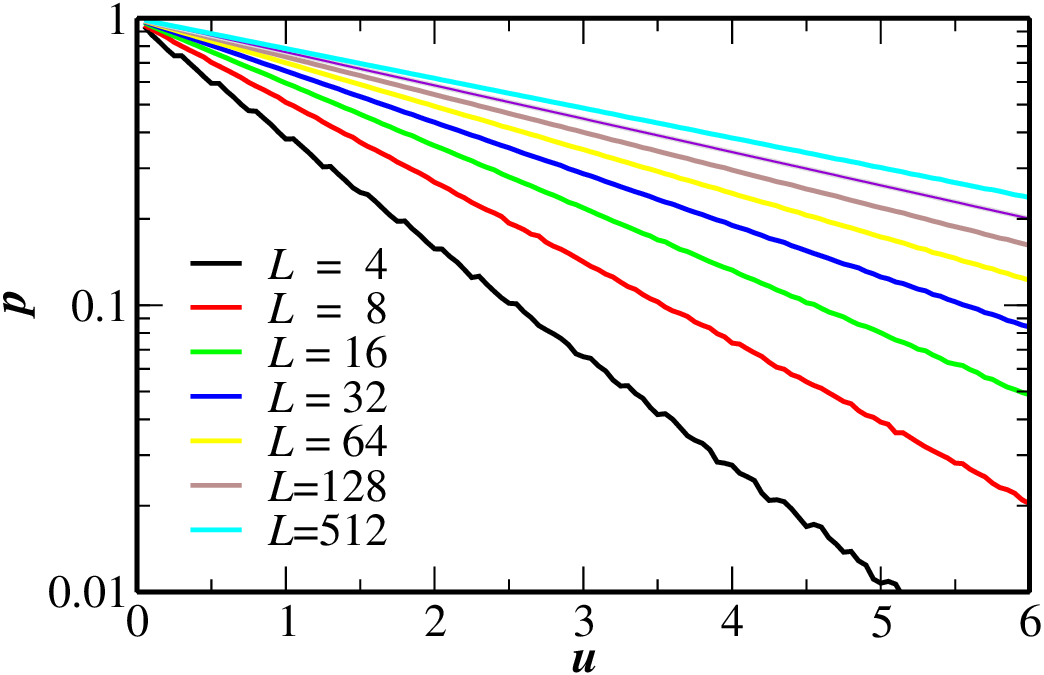}
\caption{
Semilogarithmic plot of the fraction of unvisited sites $p$ versus $u$,
for $4, 8,\dots, 512$ (bottom to top) at $d=2$.  All  graphs can be well fitted
by  $p=\exp[-A(L)u]$, where the prefactor $A(L)$ (slope on this
semilogarithmic graph), which depends on $L$,  does not saturate to a
finite value for $L\to\infty$.
}
\label{fig:2Dpu}
\end{figure}

The results of our numerical study of the dependence of $p$ on $u$ are
presented in Fig.~\ref{fig:2Dpu}. We performed simulations for
$L=4, 8, 16,\dots,512$. All  curves are well approximated by a pure
exponential, and appear as straight lines on the semi-logarithmic plots in
Fig.~\ref{fig:2Dpu}. Steps visible for $L=4$  (and, to lesser extent,
on $L=8$) are a consequence of $N$ being an {\em integer}, requiring
downward truncation to integer of $uL^2=16u$ (or $64u$). The apparent
slope $A(L)$ in Fig.~\ref{fig:2Dpu} keeps decreasing with increasing $L$.
In fact, even the product $A(L)\ln L$ has some residual dependence on
$L$, but approaches $\pi/2$ for large $L$. We verified this convergence
to five-digit accuracy. [The extra factor of 2 is due to the fact that to
leading order $\ln N=\ln(uL^2)\approx 2\ln L$.] Thus for large $L$,
\begin{equation}\label{eq:pu2D}
p=\exp\left(-\frac{\pi u}{2\ln L}\right)\ .
\end{equation}
In the asymptotic limit of large $L$, this result has been proven in
Ref.~\cite{Brummelhuis92}. The choice of $u$ over $p$ as the control
parameter is inconsequential in higher dimensions where the two
quantities are related by a fixed function. In $d=2$, the relation
between $p$ and $u$ in Eq.~\ref{eq:pu2D} depends on $L$, and using $u$
creates a somewhat different perspective.

\begin{figure}[t]
\includegraphics[width=8 truecm, clip=true]{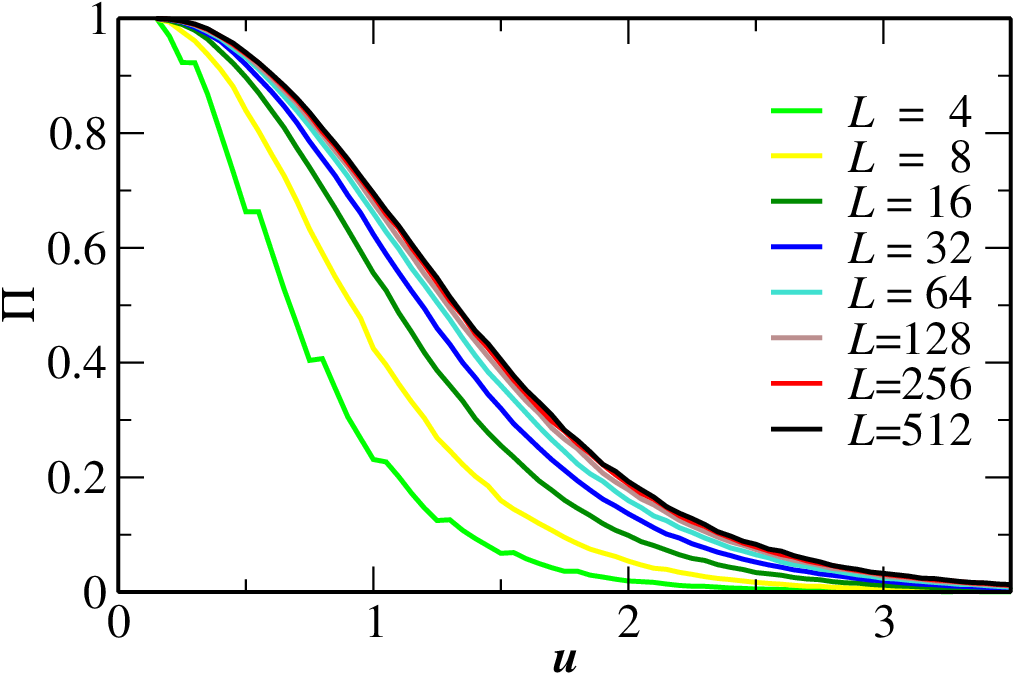}
\caption{
Spanning probability $\Pi_2(L,u)$ on an $L\times L$ lattice for random
walks of length $N=uL^2$, for $L=4,8,\dots,512$ (bottom-left to
top-right). Each point is an average of $10^5$ configurations, and the
points are separated by $\Delta u=0.05$.
}
\label{fig:2D_Pi_vs_u}
\end{figure}

Figure~\ref{fig:2D_Pi_vs_u} depicts numerically calculated accurate
values of $\Pi_2(L,u)$ obtained using large statistics ($10^5$ samples
per data point). We studied the spanning probability on a square
lattice for $L=4, 8, 16,\dots,512$; the limiting factor for the largest
$L$ was the need to  evaluate $\Pi_2(L,u)$ with high accuracy. As
before, the steps seen on the graph for $L=4$  are a result of
truncating $uL^2=16u$ to integer $N$. We immediately note that there is
no sign of $\Pi_2(L,u)$ becoming a step function with increasing $L$:
The two-dimensional problem does {\em not} have a  nontrivial
percolation  threshold. On the other hand, we clearly see that for
large $L$, the spanning probability approaches a smooth function:
$\lim_{L\to\infty}\Pi_2(L,u)=\Pi_2(\infty,u)$.

It can be argued that the existence of a finite $\Pi_2(\infty,u)$ is
a consequence of the reduced role of  lattice spacing in $d=2$.
For $d\geq 3$ the discreteness of the lattice plays a crucial role:
For percolation in the $x_d$ direction, a RW that is a one-dimensional path
must completely block passage between boundaries at $x_d=0$ and $x_d=L-1$.
As this can only be achieved by creating a $(d-1)$-dimensional continuous
{\em surface} separating the space into disconnected parts, the RW path
must either be on a lattice, or endowed with some thickness $a$ to block a
finite volume. By contrast in $d=2$, percolation, say, in the vertical
direction, can be blocked by a path  crossing the system horizontally,
which can be accomplished by a RW  of zero thickness, as long as it
acquires an extension $R\sim aN^{1/2}$, comparable to the system dimension
$aL$. The corresponding ratio $(R/L)^2$ is proportional to $u$ and
independent of $a$, would then determine the finite probability
$\Pi_2(\infty,u)$

\begin{figure}[t]
\includegraphics[width=8 truecm, clip=true]{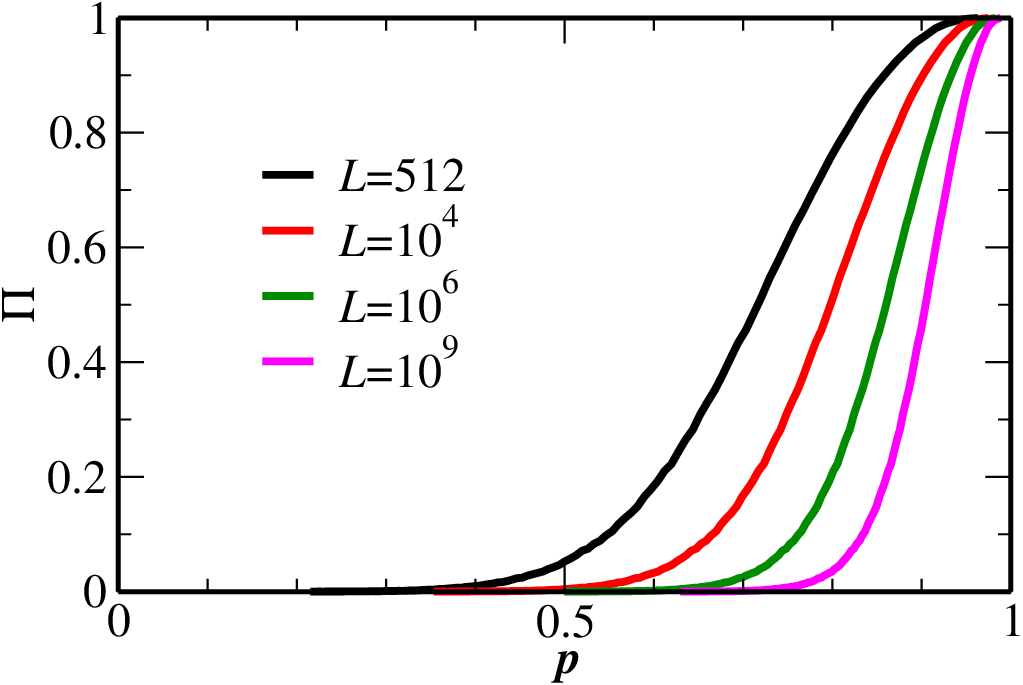}
\caption{
Semi-quantitative representation of $\Pi_2$ as a function of $p$ for
several large values of  $L$ ($L=512,10^4, 10^6$, and $10^9$, left to right).
These simulated data were built from actual data for $L=512$, while the remainder
was derived by using Eq.~\protect{\eqref{eq:pu2D}}.
}
\label{fig:2D_Pi_vs_p_semi}
\end{figure}

Since the relation between $p$ and $u$ depends on $L$, as in
Eq.~\eqref{eq:pu2D},  reexpressing $\Pi_2(\infty,u)$ as a function of $p$
leads to a function that depends on $L$. As depicted in
Fig.~\ref{fig:2D_Pi_vs_p_semi}, with increasing $L$ the corresponding
$\Pi_2(L,p)$ shift toward $p=1$. Thus $p=1$ serves as a trivial
percolation threshold in $d=2$.

\section{Very short and  long walks in $d=2$}\label{sec:ShortAndLong}
\subsection{Limit of small $u$}

\begin{figure}[t]
\includegraphics[width=6 truecm, clip=true]{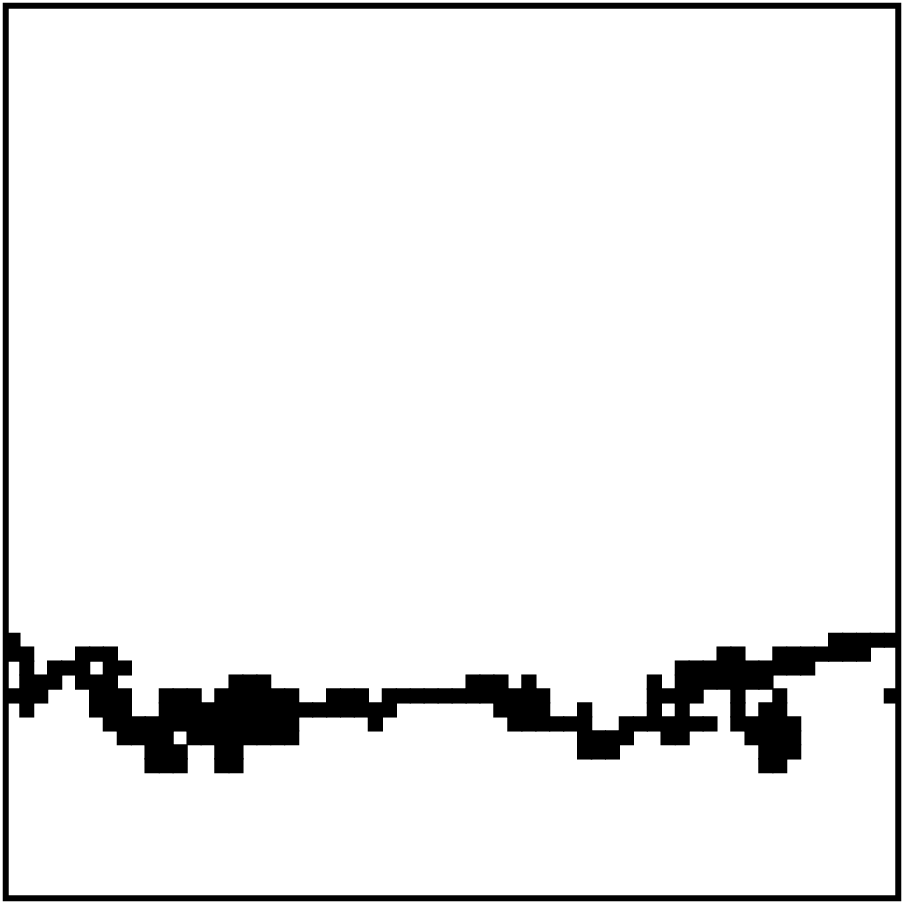}
\caption{An extremely rare event for which the vacant sites do
{\em not} percolate in the vertical direction in a $64\times 64$ lattice for
a 409-step random walk ($u=0.1$).}
\label{fig:2D_picture_u0}
\end{figure}

For $u\ll1$, the typical linear size of a RW, such as its end-to-end
distance, is $\sqrt{u}L$. Since this is shorter than the linear size,
$L$, of the lattice, a typical RW cannot block  percolation in the
vertical ($x_2$) direction, and $\Pi_2\approx1$. The only reason for
deviation of $\Pi_2$ from 1 is due to rare configurations that,
stretching far beyond the typical end-to-end distance, circumnavigate
the periodic box in the horizontal ($x_1$) direction, with end segments
intersecting the starting segments to completely block  percolation in
the vertical direction. For example, for $L=64$ at $u=0.1$ the probability of absence
of percolation is about one part per $10^5$, and Fig.~\ref{fig:2D_picture_u0}
depicts one such rare event. Since such an event requires a RW of root-mean-square
length $\sqrt{u/2}L$ in the horizontal direction to be stretched by at
least  $L$ in that direction, a lower bound for the probability of getting a
nonpercolating configuration can be obtained by integrating the Gaussian
probability distribution of the end-to-end distance from $L$ to infinity.
The resulting integral can be expressed in terms of the error function, as
\begin{equation}\label{eq:PiErf}
\Pi_2\approx{\rm Erf}(1/\sqrt{u}).
\end{equation}
This is just an estimate, since besides circumnavigating the periodic cell
horizontally, the RW must also self-intersect. Since for $u\ll1$ nonpercolating
configurations are rare, we sampled $10^6$ configurations for each data point for
small $u$ with $L=512$. Below $u=0.1$ we could not find nonpercolating
configurations. However, in the range $0.1\le u\le0.3$ we got rather
accurate values of $\Pi_2(512,u)$ that fit quite well to ${\rm Erf}(1/\sqrt{u})$,
as expected in Eq.~\eqref{eq:PiErf}.

\subsection{Limit of large $u$}

\begin{figure}[t]
\includegraphics[width=8 truecm, clip=true]{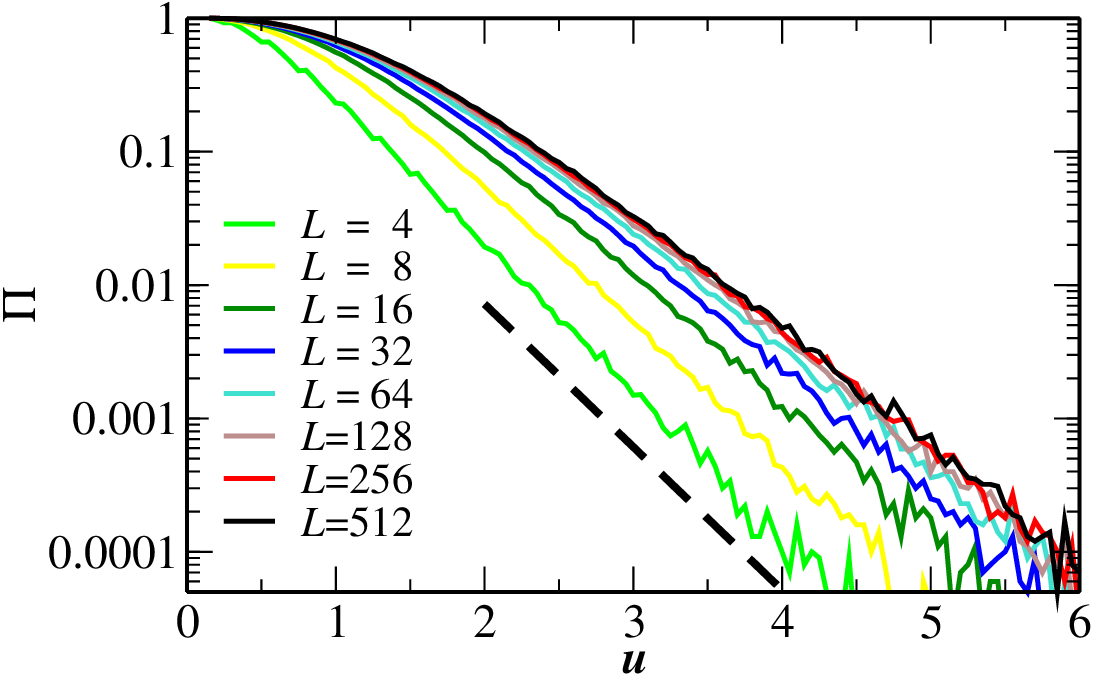}
\caption{
Same data as in Fig.~\protect{\ref{fig:2D_Pi_vs_u}} on a semilogarithmic plot.
The dashed line corresponds to $\exp(-\pi^2 u/4)$ (see text).
}
\label{fig:2D_Pi_vs_u_log}
\end{figure}

\begin{figure*}[!ht]
     \subfigure[~$u=1$\label{2D_picture_u1}]{%
       \includegraphics[width=0.4\textwidth]{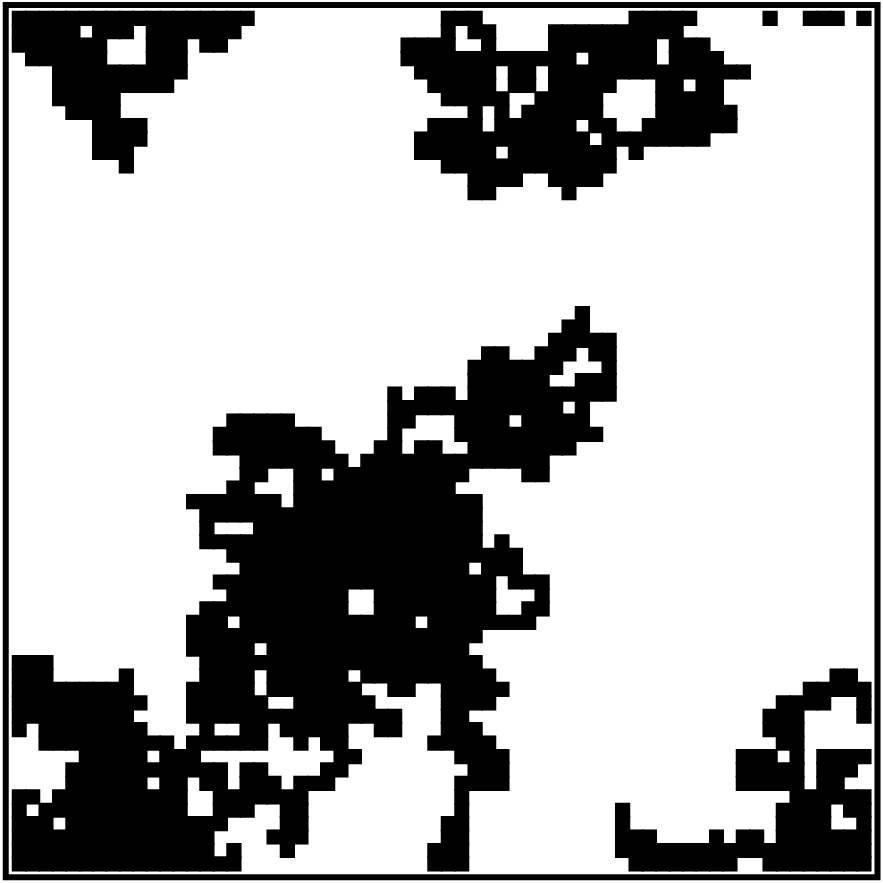}
     }
    \hfill
     \subfigure[~$u=4$\label{2D_picture_u4}]{%
       \includegraphics[width=0.4\textwidth]{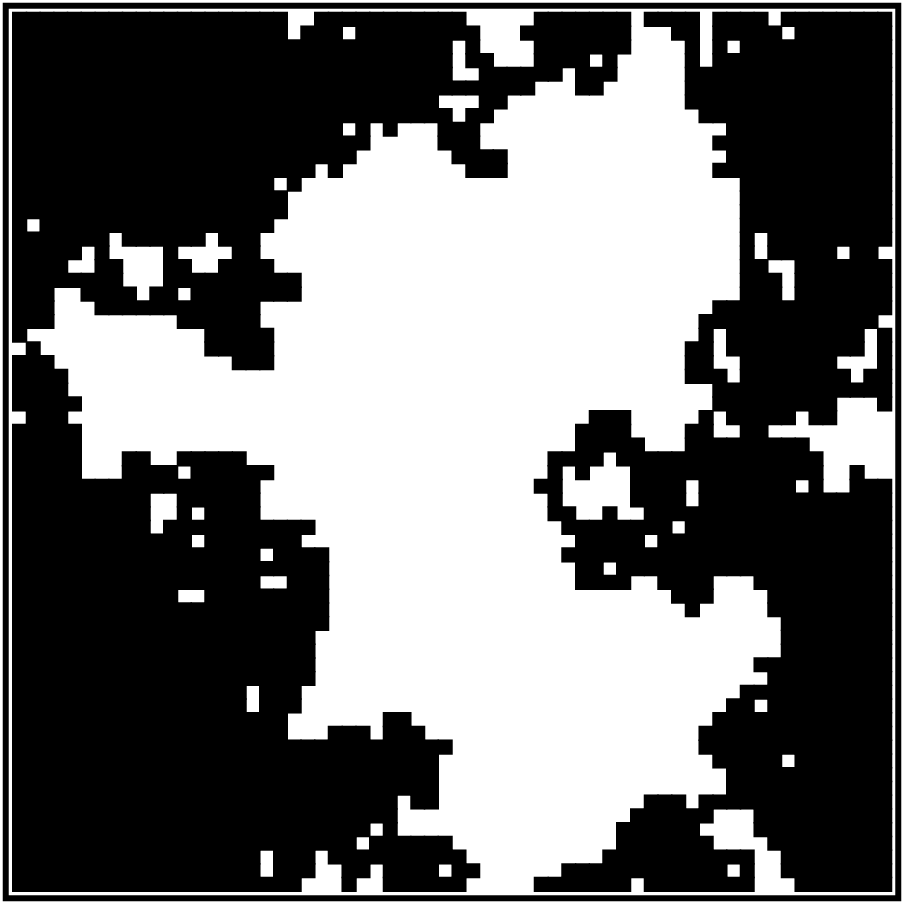}
     }
     \vskip 2cm
     \subfigure[~$u=6$\label{2D_picture_u6}]{%
       \includegraphics[width=0.4\textwidth]{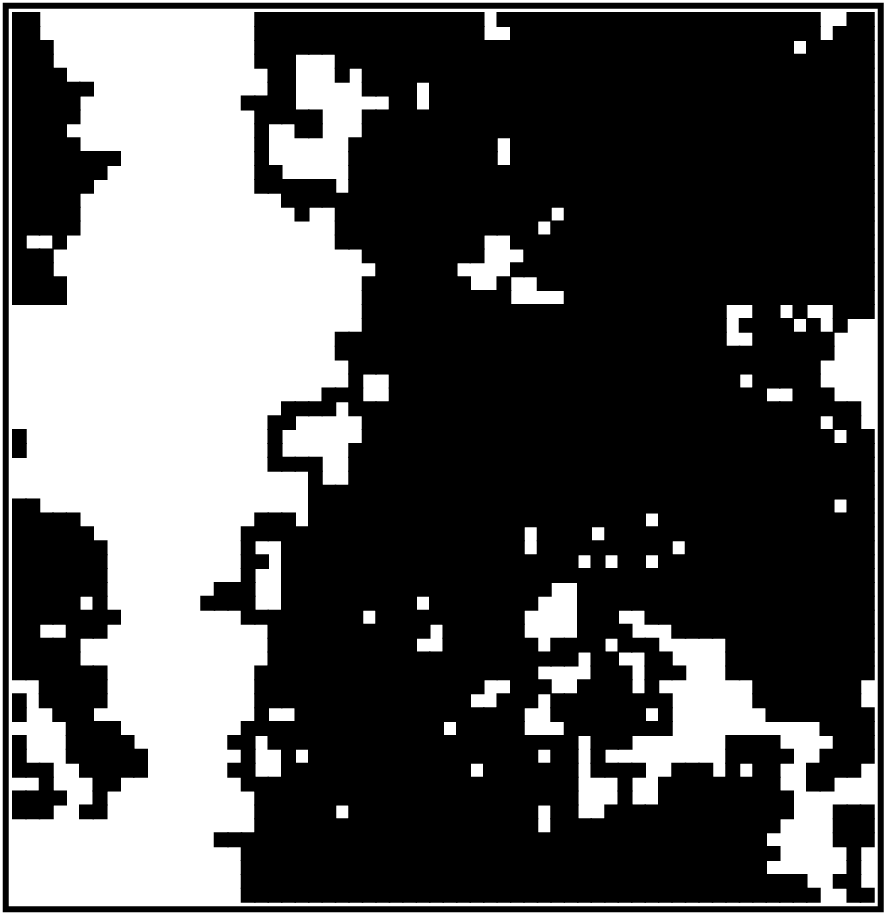}
     }
     \hfill
     \subfigure[~$u=8$\label{2D_picture_u8}]{%
       \includegraphics[width=0.4\textwidth]{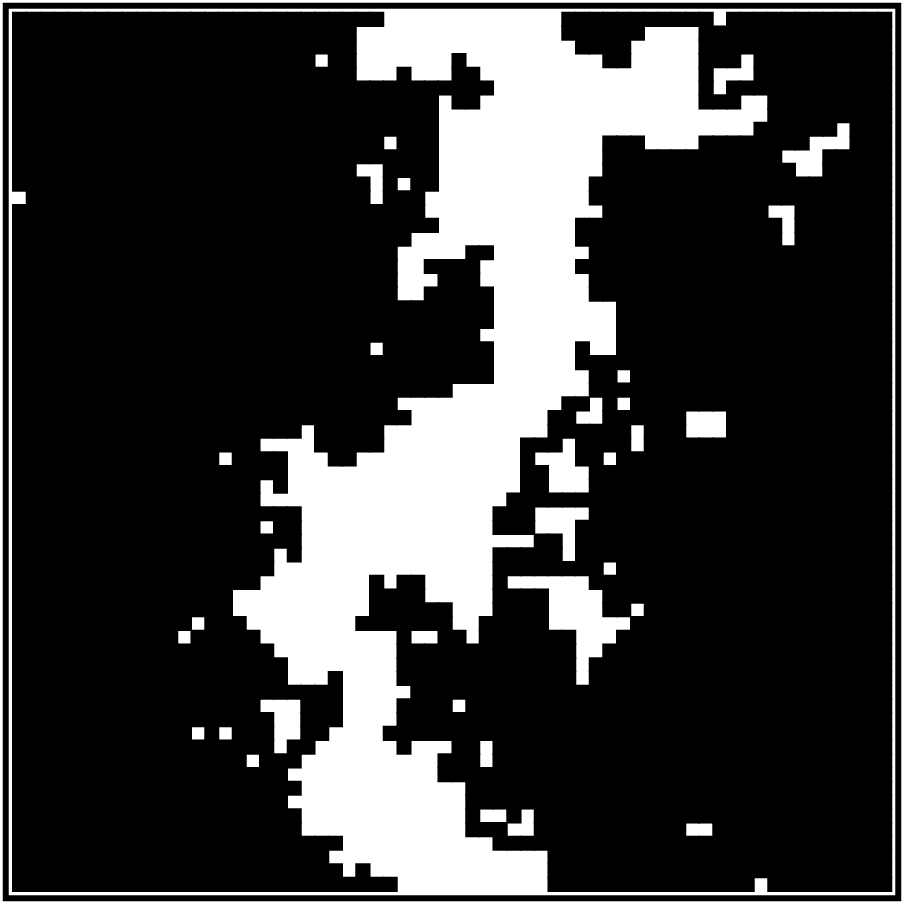}
     }
\caption{Examples of configurations on $64\times 64$ lattice, where
(white) sites unvisited by the random walk span the system in the vertical
direction.}
     \label{fig:spanning_confs}
\end{figure*}

It is interesting to view the results of Fig.~\ref{fig:2D_Pi_vs_u} on a
semilogarithmic scale, as depicted in Fig.~\ref{fig:2D_Pi_vs_u_log}.
Since each data point was obtained by averaging $10^5$ random
configurations, the statistical errors exceed 10\% for $\Pi_2$  below
$10^{-3}$, and exceed 30\% for $\Pi_2$ below $10^{-4}$ (comprising all
fluctuations in the bottom right corner of the figure). We note that
for $u>3$, the spanning probability for $\Pi_2\ll 1$ seems to decay
exponentially (while maintaining some residual dependence on $L$).
To understand this behavior of the spanning probability, we take a
closer look  at the  shapes of percolating configurations.
Figure~\ref{fig:spanning_confs} depicts four such examples, for
different values of $u$, with the black squares indicating sites
visited by the RW, while the white area corresponds to vacant sites.
By construction, the black sites always form a single cluster, while
the vacant sites can be split into many clusters. We notice that most
of the vacant sites also form a single large cluster. For $u=1$,
vacant sites percolate most of the time, and
Fig.~\ref{fig:spanning_confs}(a) represents a ``typical" configuration.
For $u=4$, 6, and 8, the percolating configurations are exceptional
(dropping to below $10^{-6}$ for $u=8$). As $u$ increases, the rare
percolating clusters start to resemble narrow white bands connecting
the top and bottom boundaries, and the probability of such a rare
configuration is estimated next.

Consider a two-dimensional RW of length $N=uL^2$ on a square
$L\times L$ lattice with periodic boundary conditions. The probability
distribution of the end point can be estimated from the diffusion
equation $\partial P/\partial N=D\nabla^2P$, with diffusion constant
$D=a^2/4$, where $a=1$ is the lattice spacing. Consider an extreme
configuration where the percolating cluster of unvisited sites is a
single straight line in the vertical direction. The allowed
configurations of RW in that case are those that can fit between two
vertical lines separated by distance $L$. The total number of $N$-step
RWs is $4^N$. The fraction of configurations that fit within a strip
of width $L$ can be found by solving the diffusion equation with
absorbing boundary conditions. The diffusion equation can be separated
into two independent parts: one for the vertical direction which
imposes no limits on the walker, and one for the horizontal component
with absorbing boundary conditions. With the horizontal coordinate
denoted by $x$, the normalized eigenstates are $\Psi_n(x,t)=\sqrt{2/L}\sin(\pi n x/L)\exp(-Dn^2\pi^2N/L^2)$.
For large $u$, and hence $N=uL^2$, the solution is dominated by the
term with  $n=1$, leading to the probability
$$P(x_0|x,t)=\frac{2}{L}\sin\left(\frac{\pi x_0}{L}\right)\sin\left(\frac{\pi x}{L}\right)
\exp(-D\pi^2N/L^2),$$
to find a random walker that started  at $x_0$ and arrived at $x$. To
obtain the total survival probability, we need to integrate over $x$
and to average over the starting point $x_0$. (The  vertical
direction is completely free and does not affect the probability.)
The final result is the  `survival probability' of
$G\propto\exp(-D\pi^2N/L^2)\propto\exp(-\pi^2u/4)$.
(We omitted the numerical prefactor of this expression.) Note that
our answer does not depend on $L$, and $G\sim \exp(-\pi^2u/4)$ provides
a lower bound for $\Pi_2$ at large $u$. This result clearly
underestimates $\Pi_2(L,u)$, as the percolating channel does not have
to be a  straight line; it can be undulating or inclined. We anticipate
that these factors will modify the prefactor for $G$, but leave
unchanged the leading exponential part that is depicted by the dashed
line in Fig.~\ref{fig:2D_Pi_vs_u_log}. It clearly underestimates $\Pi_2$
by several orders of magnitude, but its slope (on the semilogarithmic
plot) is close to the behavior of the numerical results for large $u$.

\section{Discussion}\label{sec:discussion}

In this work we studied percolation of sites unvisited by RWs on a
periodic lattice. We extended previous results by looking at larger
lattice sizes in $d=3$ and by considering all dimensions $2\le d\le6$.
Our primary goal was to find the dimensionality dependence of such
characteristics as the percolation threshold $u_c$, the critical
exponent $\nu$, and the critical spanning probability $\Pi_d^c$. While
our results agree well with the general theory of correlated percolation,
much remains unknown. We concentrated on a single critical exponent
$\nu$, and did not attempt to calculate additional critical exponents,
e.g., describing cluster sizes and fractal dimensions, which may provide
an interesting direction for further study. The RW creates rather unique
constraints on cluster structure that may lead to interesting results not
only for $d\le 6$, but also in higher dimensions when the critical behavior
is mean-field like.

The original impetus for the study of this type of percolation came from
degradation of a gel. Percolation in such a gel is accompanied by dramatic
changes in physical properties such as elasticity. However, theoretical
studies (with the exception of Ref.~\cite{Banavar85}) have so far concentrated
on purely geometrical quantities. Thus it would be interesting to extend the
theoretical investigation of such correlated percolation to conductivity and
diffusion.

\begin{acknowledgments}
Y.K.~thanks D. Gomez for stimulating discussions on related subjects.
M.K.~was supported by the National Science Foundation through Grant No.~DMR-1708280.
Y.K.~was supported by the  Israel Science Foundation Grant No.~453/17.
\end{acknowledgments}

\end{document}